\documentclass[lettersize,journal]{IEEEtran}
\usepackage{amsmath,amsfonts}
\usepackage{amssymb}
\usepackage{algorithmic}
\usepackage{algorithm}
\usepackage{array}
\usepackage[caption=false,font=normalsize,labelfont=sf,textfont=sf]{subfig}
\usepackage{textcomp}
\usepackage{stfloats}
\usepackage{CJKutf8}
\usepackage{url}
\usepackage{verbatim}
\usepackage{graphicx}
\usepackage{cite}
\usepackage{caption}

\def\BibTeX{{\rm B\kern-.05em{\sc i\kern-.025em b}\kern-.08em
    T\kern-.1667em\lower.7ex\hbox{E}\kern-.125emX}}

\begin{document}
\begin{CJK}{UTF8}{gbsn}

\title{Exploring Neurofunctional Phase Transition Patterns in Autism Spectrum Disorder: A Thermodynamics Parameters Analysis Approach}

\author{Dayu Qin, \IEEEmembership{Student Member, IEEE}, Yuzhe Chen, and Ercan Engin Kuruoglu, \IEEEmembership{Senior Member, IEEE}
\thanks{Dayu Qin, Yuzhe Chen, and Ercan Engin Kuruoglu are with the Tsinghua-Berkeley Shenzhen Institute, Tsinghua Shenzhen International Graduate School, Tsinghua University
(e-mail: tdy22mails.tsinghua.edu.cn; chen-yz22@mails.tsinghua.edu.cn; kuruoglu@sz.tsinghua.edu.cn).}
\thanks{Ercan Engin Kuruoglu is the corresponding author.}
}

% The paper headers
\markboth{Arxiv preprint}{Qin \MakeLowercase{\textit{et al.}}: Exploring Neurofunctional Phase Transition Patterns in Autism Spectrum Disorder}

\maketitle

\begin{abstract}
Designing network parameters that can effectively represent complex networks is of significant importance for the analysis of time-varying complex networks. This paper introduces a novel thermodynamic framework for analyzing complex networks, focusing on Spectral Core Entropy (SCE), Node Energy, internal energy and temperature to measure structural changes in dynamic complex network. This framework provides a quantitative representation of network characteristics, capturing time-varying structural changes. We apply this framework to study brain activity in autism versus control subjects, illustrating its potential to identify significant structural changes and brain state transitions. By treating brain networks as thermodynamic systems, important parameters such as node energy and temperature are derived to depict brain activities. Our research has found that in our designed framework the thermodynamic parameter—temperature, is significantly correlated with the transitions of brain states. Statistical tests confirm the effectiveness of our approach. Moreover, our study demonstrates that node energy effectively captures the activity levels of brain regions and reveals biologically proven differences between autism patients and controls, offering a powerful tool for exploring the characteristics of complex networks in various applications.
\end{abstract}


\section{Introduction}
\label{introduction}
\IEEEPARstart{H}{uman} brains' resting state activity cannot be described as random. Instead, there are governing rules of brain activity at different time scales, ranging from a few seconds to even several years (developmental). It has been discovered that resting-state functional connectivity has several cognitive phenotypes \cite{gonzalez2018task} and that spontaneous brain activities are highly organized into hierarchical structures.
This brings forward the study of brain states and brain state transitions. As defined in \cite{greene2023everyone}, brain states are recurring brain configurations emerging from physiological or cognitive processes. It is a building block of neural dynamic models, advancing the understanding of brain activities. The study of brain state transition can facilitate Sleep stage detection \cite{stevner2019discovery}, and the differences between brain state compositions can distinguish task and rest brain activities \cite{gonzalez2018task}. The brain state occupancy profile can serve as identifiers of individual differences and biomarkers of Neurological disorders (ADHD \cite{yamashita2021brain}, schizophrenia \cite{yu2015assessing}, major depressive disorder \cite{javaheripour2023altered})

However, it is hard to detect and quantify the transition of brain states, as the concept of a brain state is elusive and incoherent in different fields of research. Also, as the temporal resolution of brain signals varies with different modalities and the target scale changes for different research, the temporal scale of observable brain activities will impact brain-state related definitions and findings \cite{hari2015brain}. 

In current studies, there are several ways to categorize brain signals into brain states. Brain signals can be categorized according to biological definition (e.g. brain wave, phase synchronization, potential distribution). Based on different functional networks in the brain, previous work defines brain states to be composed of different sub-states and can be linear combinations of them \cite{stevner2019discovery}. However, such definitions are not generalizable, as the biological definition can be flawed, which can limit the emergence of novel findings. 

There are also data-driven definitions of brain states. There are studies that perform classification or clustering to identify groups of similar brain configurations. Critical findings have also been made by using Hidden Markov Models (HMMs) \cite{vidaurre2017brain, stevner2019discovery}. However, there are also clear disadvantages of clustering algorithms. For instance, the K-Means algorithm requires the number of clusters to be known or defined beforehand. For algorithms such as DBSCAN, not all data samples (time instants) are clustered. Previous studies have also used matrix distance to quantify the change in brain connectivity networks \cite{lee2019brain}. However, current metrics cannot reflect the brain network connectivity and detect the change in matrix structure, which is an important trait of brain activities. 

Graphs are powerful tools for dynamic brain connectivity analysis since they are suitable for representing co-functioning activities in the brain. Graph signal processing (GSP) further enhances the analysis of network data. GSP methods have been devised to detect robust graph signals against noise \cite{li2023robust} and boost computational efficiency \cite{yan2022adaptive}, adding potential to this field. Previously, GSP has also been used to analyze time-varying financial networks and financial market crash prediction \cite{qingraph}. GSP has a wide range of applications for brain network analysis. Huang et al. \cite{huang2016graph} decompose brain signals according to different smoothness (or rapidness) levels by using graph spectral operations and discovered the correspondence between brain network activities and graph frequencies. 
Preti et al. \cite{preti2023graph} leverage the structure-function coupling implication of graph frequency components, linking brain functional activities with neural architectures through node- and edge-level graph metrics. However, these methods construct graphs based on brain regions while the time information is collapsed in the correlation calculation step, hindering the detailed analysis of brain
state distributions across time. However, there is a lack of an effective summary indicator that helps depict the change in graph structures and transitions in brain states. 

Given the research gap mentioned above, a quantitative representation that better captures the change in brain connectivity structure is needed to identify brain state transitions. We introduce a thermodynamic framework to measure the structural change in dynamic brain connectivity.

In this paper, we define a brain state as a category of similar instantaneous connectivity patterns. We also assume that the change in brain connectivity should be continuous within each brain state, but we expect drastic transitions between brain states. Hence, we will aim to detect the transition points, giving us a continuous brain state in the interval between two transition points. To evaluate the effectiveness of this new metric, we apply it to resting-state functional magnetic resonance imaging (fMRI) data to study brain state transitions for autism versus control subjects.

The concept of entropy proposed within the framework of complex networks is commonly considered a measure of the network's complexity \cite{ye2015thermodynamics}. In most definitions, the calculation of entropy is closely related to the Laplacian matrix of the network. However, for networks with a large number of nodes and edges, the Laplacian matrix is not easy to compute. Therefore, if we can approximate the entropy calculation without compromising the representation of the network's structural characteristics, it will significantly save computational resources.

Our paper will consist of the following sections: The first section will serve as an introduction and include some literature review. The second section will introduce our proposed novel Spectral Core Entropy and the corresponding thermodynamic framework. The third section will present our experimental results. In the fourth section, we will analyze and discuss our results. The fifth section will provide conclusions and discuss future research directions.

\section{Thermodynamics Framework for Graph Data}\label{thermo}
In this section, we introduce the innovative concepts of Spectral Core Entropy and Node Energy, which allow us to derive the internal energy and temperature of a graph. This leads to the development of a novel thermodynamic framework for complex networks.

\subsection{Preliminary Constructions}

For an undirected graph $G(V,E)$ with node set $V$ and edge set $E$, its adjacency matrix $A$ can be defined as following:

\begin{equation}
    A_{i j}= \begin{cases}1, & (i, j) \in E \\ 0, & \text { else }\end{cases}
\end{equation}

\noindent where $D$, its degree matrix, is a diagonal matrix and can be expressed as:
\begin{equation}
    D_{ik}=\begin{cases}\sum_j A_{i j}, & i = k  \\ 0, & \text { else }\end{cases}
\end{equation}

The Laplacian matrix $L$ of the graph is the difference between the degree and adjacency matrices, i.e., $L = D-A$. $L$ can be further normalized, leading to the normalized Laplacian matrix $\tilde{L}=D^{-1 / 2}(D-A) D^{-1 / 2}$. Each element in the matrix can be expressed as:
\begin{equation}
    \tilde{L}_{i j}=\left\{\begin{array}{cl}
    1 & \text { if } i=j \text { and } d_i \neq 0 \\
    -\frac{1}{\sqrt{d_i d_j }} & \text { if } i \neq j \text { and }(i, j) \in E \\
    0 & \text { otherwise. }
    \end{array}\right.
\end{equation}

 The eigenvalues ${\lambda}_i$ of the normalized Laplacian matrix have the property $\sum_{i=1}^{|V|} \tilde{\lambda}_i = \operatorname{Tr}[\tilde{L}]=|V|,  i=1,2, \ldots,|V|$.

In our work, we consider the complex network to behave as a thermodynamics system. That is to say, we assume the system will occupy $V$ microstates with a certain probability distribution. The probability of the system occupying a microstate will be $p_i=\tilde{\lambda}_i / \sum_{s=1}^{|V|} \tilde{\lambda}_i = \tilde{\lambda}_i /|V|$, where $\tilde{\lambda}_i, i=1,2, \ldots,|V|$ represent the eigenvalues of the normalized Laplacian matrix.

Then the Boltzmann entropy of the system can be written as the following:
\begin{equation}
S = -k_B \sum_i^N p_i \ln p_i=- k_B \sum_i^N \frac{\tilde{\lambda}_i}{|V|} \ln \frac{\tilde{\lambda}_i}{|V|}
\end{equation}

\noindent where $k_B$ is the Boltzmann constant.

\subsection{Spectral Core Entropy (SCE)}
In previous work \cite{ye2018thermodynamic}, the entropy is approximated using Taylor expansion of $ln(x)$ at $x = 1$. The simplified form of $-\ln \frac{\tilde{\lambda}_i}{|V|}$, which is $(1-\frac{\tilde{\lambda}_i}{|V|})$, has significant advantages when calculating the upper and lower bounds of entropy. However, from the microstate probability $p_i = \tilde{\lambda}_i /|V|$, we can see that when $|V|$ is large, the value of $p_i$ tends to be closer to 0. This is also observed in practical experiments when there are over hundreds of nodes. Therefore, we consider this approximation to be not very appropriate. It is thus necessary to propose a more accurate approximation method that better reflects reality.

For a complex network that can be represented by a graph, we consider the eigenvalues of its Laplacian matrix to represent signal frequencies in the frequency domain. In the context of a thermodynamic system, these eigenvalues correspond to the microstates that contribute to the system's complexity.

For large-scale networks, calculating all eigenvalues is often impractical or unnecessary. If only the primary states within the thermodynamic system are considered and we ignore the Boltzmann constant, an approximate value of entropy, the Spectral Core Entropy can be designed as the following:

\begin{equation}
S = -\sum_i^{m} p_i \ln p_i=- \sum_i^{m} \frac{\tilde{\lambda}_i}{|V|} \ln \frac{\tilde{\lambda}_i}{|V|}.
\end{equation}

\noindent where $m$ represents the top $m$ largest eigenvalues and is an adjustable hyperparameter.

Since the Laplacian matrix is a semi-positive definite symmetric matrix, we can utilize the Lanczos algorithm to solve for the top $m$ largest eigenvalues. The efficiency of the Lanczos algorithm is higher than performing eigenvalue decomposition directly on the entire matrix, especially when dealing with large-scale problems\cite{koch20118}.

By focusing on the top $m$ largest eigenvalues, the preeminent time-varying trends of the system's physical quantities can be effectively captured, while significantly reducing computational complexity and time. In a real physical system, this might correspond to the predominant energy modes, the strongest interactions, or the principal determinants of system stability. It is because in a thermodynamics system, the entropy of a system is related to the weight of its microscopic states. The largest entropy components represent the states with the highest statistical weight, indicating that the most probable energy modes in the system. Moreover, the interactions between particles within the system determine its energy. The largest entropy components are often associated with the strongest interactions.

\subsection{State Energy, Internal Energy, and Temperature}
\subsubsection{Node Energy}
We have obtained an approximation of the system's entropy. However, to establish a complete thermodynamic structure, we also need to define the system's internal energy and temperature. In our design, we assume that nodes, similar to particles, possess energy and represent a part of the matter in a physical system. The some certain changes of nodes indicate the changes of the energy of the particles over time. By treating nodes as particles within the thermodynamic system, we define the energy of the nodes in the following form:

\begin{equation}
U_i=\sum_j \frac{1}{d_i+d_j} d_i \varepsilon_{i j}
\end{equation}

\noindent$d_i$ is the degree of node i, j is the set of neighbour nodes of node i and $\varepsilon_{i j}$ is the edge between node i and j. Since we use adjacency matrix, $\varepsilon_{i j} = 1$.

We consider edges as total energy of the thermodynamics system, which is internal energy, and in the later derivation, we can see that this assumption is self-consistent. The particles in the system distribute this energy according to the ratio of their own charge, their node degree, to the charge of their first-order neighboring nodes. According to the Principle of locality\cite{tong2006lectures}, we aim to define the node energy in this way to establish a simple form of energy which only focus on the connection between a node and its neighboring nodes. It is similar to the simplified calculation of intermolecular forces for network-like structure material that we can ignore the influence other than the particles' nearest neighbors because the intermolecular forces decays rapidly with distance. Since the state energy of the system is correlated with the degree of the node, we name it Node Energy.

\subsubsection{Internal Energy}
If we consider our system an isolated system, then the network internal energy will be only contributed by particles' energy, which means we only need to add up all the nodes' energy to get the internal energy. Thus, the expression of internal energy can be written as: 

\begin{equation}
U=\sum_i^N U_i=\sum_i^N   \sum_j \frac{1}{d_i+d_j} d_i \varepsilon_{i j} =|E|.
\end{equation}

In our framework, the internal energy of the system is given by the total number of edges in the graph. This derived result aligns with our previous assumption that each edge represents a unit of energy. That is to say, in the thermodynamic framework we constructed, the total number of nodes does not directly impact the total internal energy of the system. The designed energy focuses more on the connections between nodes—edges. However, the degree information of the nodes is included in the Node energy, which can serve as a node characteristic. This can help us analyze specific node behaviors within this overall framework.

\subsubsection{Temperature}
Since we consider the graph as an isolated thermodynamic system, there is a certain relationship between entropy, internal energy, and temperature \cite{callen1991thermodynamics} :
\begin{equation}
d U \leqslant T d S+\delta w
\end{equation}

Since we consider this to be a reversible process and there is no external work involved, the above equation can be written as follows:

\begin{equation}
d U = T d S
\end{equation}

That is,
\begin{equation}
T=\frac{\Delta U}{\Delta S}
\end{equation}

In our framework, for two consecutive time points $t = n-1$ and $t = n$, the temperature can be represented in the following discrete form:

\begin{equation}
T_n=\left\{\begin{array}{l}
0, n=1 \\
\frac{U_n-U_{n-1}}{S_n-S_{n-1}}, n \geqslant 2
\end{array}\right.
\end{equation}

\noindent$U_n$, $U_{n-1}$, $S_n$, and $S_{n-1}$ are the values of internal energy and spectral core entropy at $t=n$ and $t=n-1$, respectively.

From this point on, for each time point's graph, we can calculate the network's SCE, each node's energy, internal energy, and temperature. This completes the thermodynamic framework that can be used to characterize the time-varying graph. 

Next, we will observe the practical utility of our designed thermodynamic framework in our experiments. These results demonstrate that the behavior of our designed thermodynamic parameters in complex networks aligns with theoretical expectations, potentially revealing the applicability of our thermodynamic framework to graph learning.

\section{Brain signal analysis procedure}
\begin{figure}[t!]
    \centering
    \includegraphics[width=0.4\textwidth]{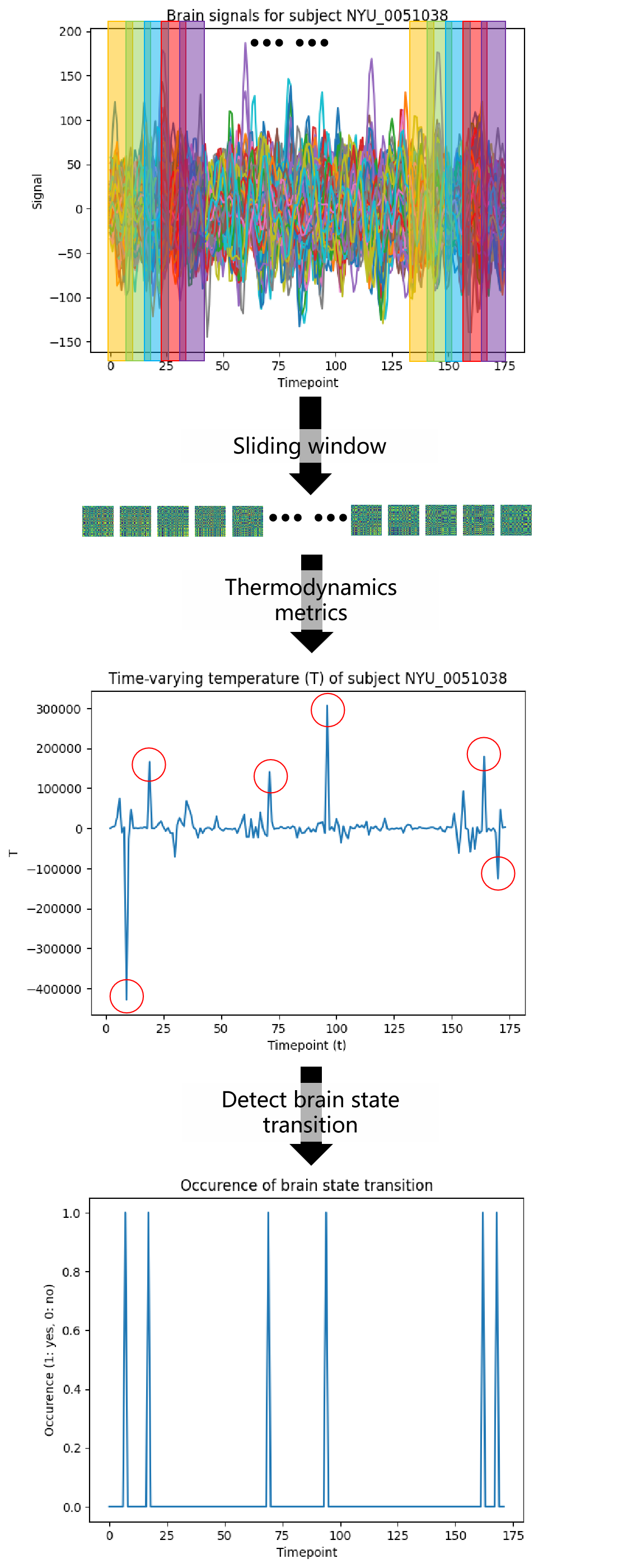}
    \caption{Algorithm diagram: Brain connectivity matrices are constructed using a sliding window approach. Thermodynamics metrics are then applied to extract the time-varying structural changes and detect brain state transition points.}
    \label{fig:algo_diag}
\end{figure}

With the mathematical formulation of the metrics ready, we can perform the following analysis procedure to study brain signals, including sliding window, thermodynamic calculation, and statistical correlation testing, as Fig. \ref{fig:algo_diag} shows.

First, we adopt a sliding window approach that divides the time courses into overlapping time intervals, in order to extract finer-grained time-varying brain connectivity. Here, we describe the mathematical formulation of the sliding window approach.

\begin{itemize}
    \item \textbf{Windowing:} Given an fMRI time series data matrix $\mathbf{X} \in \mathbb{R}^{T \times N}$, where $T$ represents the number of time points and $N$ denotes the number of brain regions (or nodes), we segment the data into overlapping windows. Let $w$ be the window length and $s$ be the step size between consecutive windows. The $k$-th windowed segment of the time series data, denoted as $\mathbf{X}_k$, is defined as:
\begin{equation}
    \mathbf{X}_k = \mathbf{X}[t_k:t_k+w-1, :]
\end{equation}

where $t_k = k \cdot s$ and $k \in \{0, 1, \ldots, \left\lfloor \frac{T-w}{s} \right\rfloor\}$.

\item \textbf{Connectivity Matrix Calculation}: 
For each windowed segment $\mathbf{X}_k$, we compute the connectivity matrix $\mathbf{A}_k \in \mathbb{R}^{N \times N}$. The connectivity between regions $i$ and $j$ within the $k$-th window can be calculated using Pearson correlation. We set the diagonal elements to zero to rule out self-connection. In this way, we formulate the connectivity (adjacency) matrix as:

\begin{equation}
    \mathbf{A}_k(i,j) = \begin{cases}
        \frac{\text{Cov}(\mathbf{X}_k(:,i), \mathbf{X}_k(:,j))}{\sigma(\mathbf{X}_k(:,i)) \cdot \sigma(\mathbf{X}_k(:,j))}, & i \neq j \\ 0, & i = j
    \end{cases}
\end{equation}

\noindent where $\text{Cov}(\mathbf{X}_k(:,i), \mathbf{X}_k(:,j))$ is the covariance between the time series of regions $i$ and $j$ within the $k$-th window, and $\sigma(\mathbf{X}_k(:,i))$ and $\sigma(\mathbf{X}_k(:,j))$ are the standard deviations of these time series.

\end{itemize}

Second, the adjacency matrices are calculated within each window as the correlation between the signals of two brain regions, thus giving us a time-varying brain network. 

Third, we summarize each matrix's structural information using thermodynamic metrics, calculated as defined in \ref{thermo}. As \ref{fig:algo_diag} illustrates, spike values exist at some time points. We set a threshold to select the extreme values, as they indicate significant structural changes in connectivity matrices and hence reveal the time instants where brain state transitions occur.

To verify the effectiveness of the proposed detection method, we adopt the structural similarity (SSIM) metric well established in image processing as the baseline for comparison. The SSIM index measures the similarity between two images. As formulated in Eqs.~\eqref{eq: SSIM}-\eqref{eq: SSIM2}, SSIM is composed of brightness, contrast, and structure measures. Viewing the adjacency matrices as 2-dimensional grey images, the brightness reflects their overall connectivity level, the contrast depicts the variance of connectivity strength among brain regions, and the structural information captures the configuration of the brain.

The formula for the SSIM index between two images \( x \) and \( y \) is defined as follows:

\begin{equation}
\text{SSIM}(x, y) = \frac{(2\mu_x \mu_y + C_1)(2\sigma_{xy} + C_2)}{(\mu_x^2 + \mu_y^2 + C_1)(\sigma_x^2 + \sigma_y^2 + C_2)}
\label{eq: SSIM}
\end{equation}

\noindent where \( \mu_x \) and \( \mu_y \) represent the mean values of images \( x \) and \( y \), respectively, \( \sigma_x^2 \) and \( \sigma_y^2 \) denote the variances of \( x \) and \( y \), and \( \sigma_{xy} \) is the covariance of \( x \) and \( y \). The constants \( C_1 \) and \( C_2 \) are introduced to stabilize the division when the denominator is very small and are typically defined as \( C_1 = (K_1 L)^2 \) and \( C_2 = (K_2 L)^2 \), where \( L \) is the dynamic range of the pixel values, and \( K_1 \) and \( K_2 \) are small constants (e.g., \( K_1 = 0.01 \) and \( K_2 = 0.03 \)).

The SSIM index can be decomposed into three multiplicative terms representing luminance, contrast, and structure. These are given by:

1. Luminance: This component measures the overall connectivity level of the brain network. High luminance indicates strong average connectivity across the brain regions.

2. Contrast: This component reflects the variance in connectivity strengths among brain regions. Contrast allows us to observe how brain connectivity changes over time. By analyzing the contrast within sliding windows of fMRI data, we can detect periods of high variability, which may correspond to significant cognitive or behavioral events\cite{leming2021single}.
.

3. Structure: This component the configuration of the brain's connectivity patterns. Structural similarity reveals how the organization of the brain network evolves over time or under different conditions.
%measures the similarity in structural information between the two images. The covariance \( \sigma_{xy} \) represents how the pixel values of one image vary relative to the other. The structure term compares the covariance with the product of the standard deviations, effectively capturing the similarity in the underlying structures of the images.

Combining these components, the SSIM index is expressed as:
\begin{equation}
\text{SSIM}(x, y) = [l(x, y)]^\alpha \cdot [c(x, y)]^\beta \cdot [s(x, y)]^\gamma
\label{eq: SSIM2}
\end{equation}

\noindent where typically \( \alpha = \beta = \gamma = 1 \), leading to the simplified formula presented initially. The SSIM index thus combines these three components to produce a single value that represents the overall similarity between the two images, where a value of 1 indicates perfect structural similarity.

We calculate the correlation between the occurrences of temperature spikes and low SSIM values using Matthew's Correlation Coefficient (MCC), which is designed for binary (occur or not) vectors. We then run a permutation test to determine the likelihood of obtaining a correlation that is at least as extreme, so as to determine the statistical significance of this correlation. 

\section{Experiment}

In the experiments, we examine brain network activity using the thermodynamics metrics proposed in Section \ref{thermo}. The experiments aim to exemplify the underlying meanings of these thermodynamic metrics in neuroscience scenarios. 

\subsection{Data acquisition}
The experiment data includes the study of time-varying brain connectivity of autism and normal people. By using the Nilearn package \cite{abraham2014machine} in Python, the pre-processed data in the ABIDE dataset \cite{craddock2013neuro} can be obtained. The pre-processed dataset version with bandpass filtering, global signal regression, and quality check is available in the data repository, ensuring a higher data quality. In order to extract time series from fMRI data, a further step of registration and masking using a template or atlas is still needed. In this experiment, the template icbm152 is used. These pre-processing steps are crucial in brain data analysis, as factors like individual differences and motion noise can lead to data ambiguity.

\subsection{Time varying thermodynamics parameter curves}
Using a sliding window approach, we can construct the time-varying brain network. For the ABIDE dataset, subjects are typically scanned for a few minutes with around 3-second temporal resolutions. We hence select the window length to be 5 time instants and the stride to be 1 time instant, in order to observe finer dynamics in the scale of seconds. In order to capture the predominant interactions and reduce computational complexity, we select the 20 largest eigenvalues ($\alpha=20$) for SCE computation. Then, we can calculate the thermodynamic parameter value under each window frame. 

\begin{figure*}[t!]
    \centering
    \subfloat[Von Neumann entropy - control]{\label{fig:e_control}
       \includegraphics[width=0.45\textwidth]{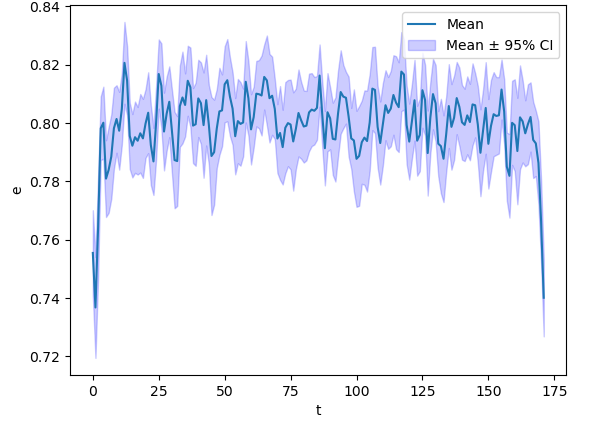}}
    \hfill
    \subfloat[Von Neumann entropy - autism]{\label{fig:e_autism}
       \includegraphics[width=0.45\textwidth]{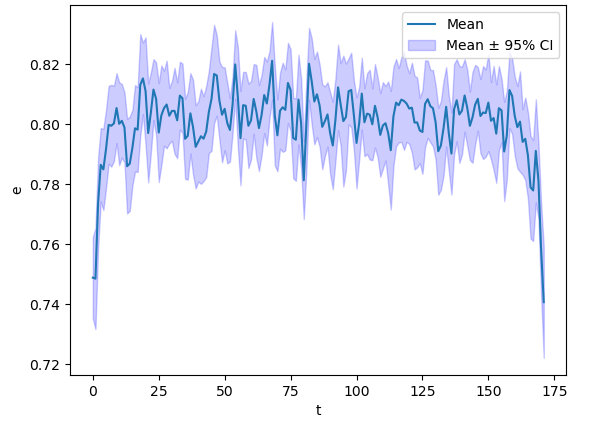}}\\
    \subfloat[Internal energy - control]{\label{fig:u_control}
       \includegraphics[width=0.45\textwidth]{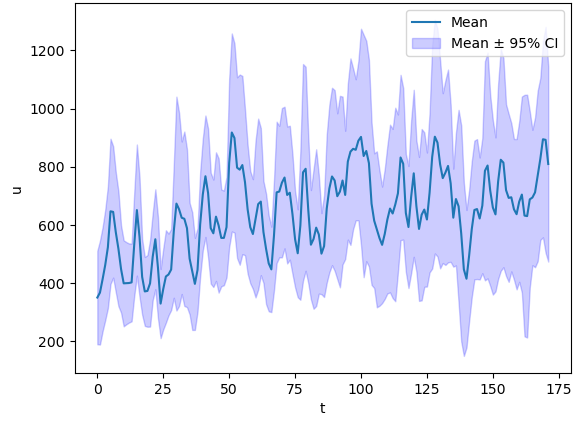}}
    \hfill
    \subfloat[Internal energy - autism]{\label{fig:u_autism}
       \includegraphics[width=0.45\textwidth]{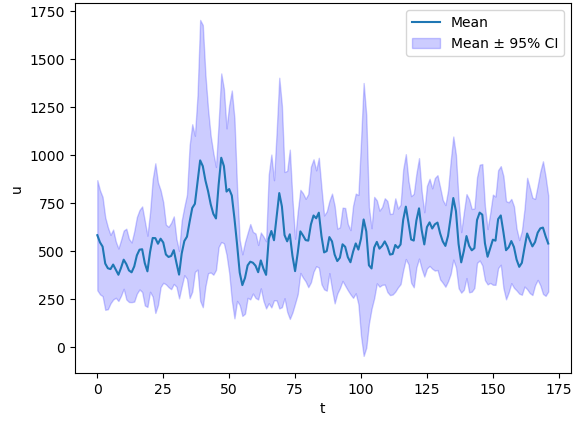}}\\
    \subfloat[Temperature - control]{\label{fig:t_control}
       \includegraphics[width=0.43\textwidth]{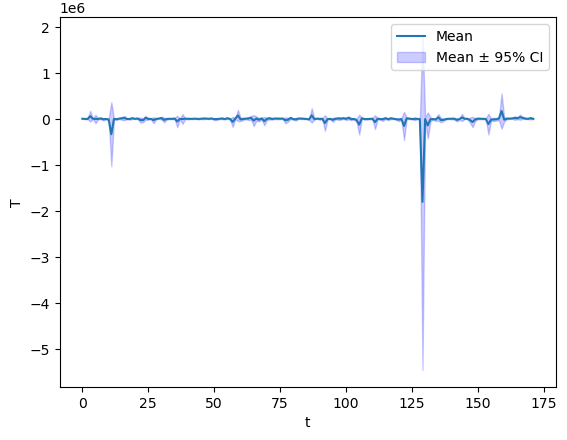}}
    \hfill
    \subfloat[Temperature - autism]{\label{fig:t_autism}
       \includegraphics[width=0.45\textwidth]{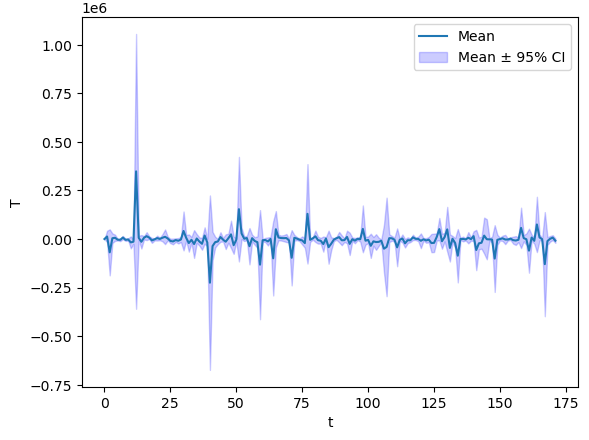}}
    \caption{Curves of time-varying entropy, internal energy, and temperature for normal and autism groups.}
    \label{curves}
\end{figure*}

On the group level, we compute the mean value and 95\% confidence interval (CI) and plot the curves for entropy, internal energy, and temperature in Fig.\ref{curves}. Both control and autism subjects experience sharp fluctuations in entropy values, indicating that the brains experience a wide variety of different configurations over time. As for the internal energy plots, the autism subjects exhibit more pronounced peaks and wider confidence intervals, suggesting greater variability and possibly more erratic changes in internal energy. The control subjects, while also showing variability, appear to have more stable internal energy with less extreme peaks and narrower confidence intervals. On the temperature plots, the autism subjects show more fluctuations with significant spikes in temperature. The control subjects, however, have more consistent temperature measurements with less extreme anomalies. These differences may indicate distinct patterns of internal energy and temperature dynamics between autism and control subjects, which could be relevant for understanding the underlying neurofunctional mechanisms.

\begin{figure*}[h!]
    \centering
    \subfloat[Scatter\_e\_u\_T\_control]{\label{fig:a}
       \includegraphics[width=0.45\linewidth]{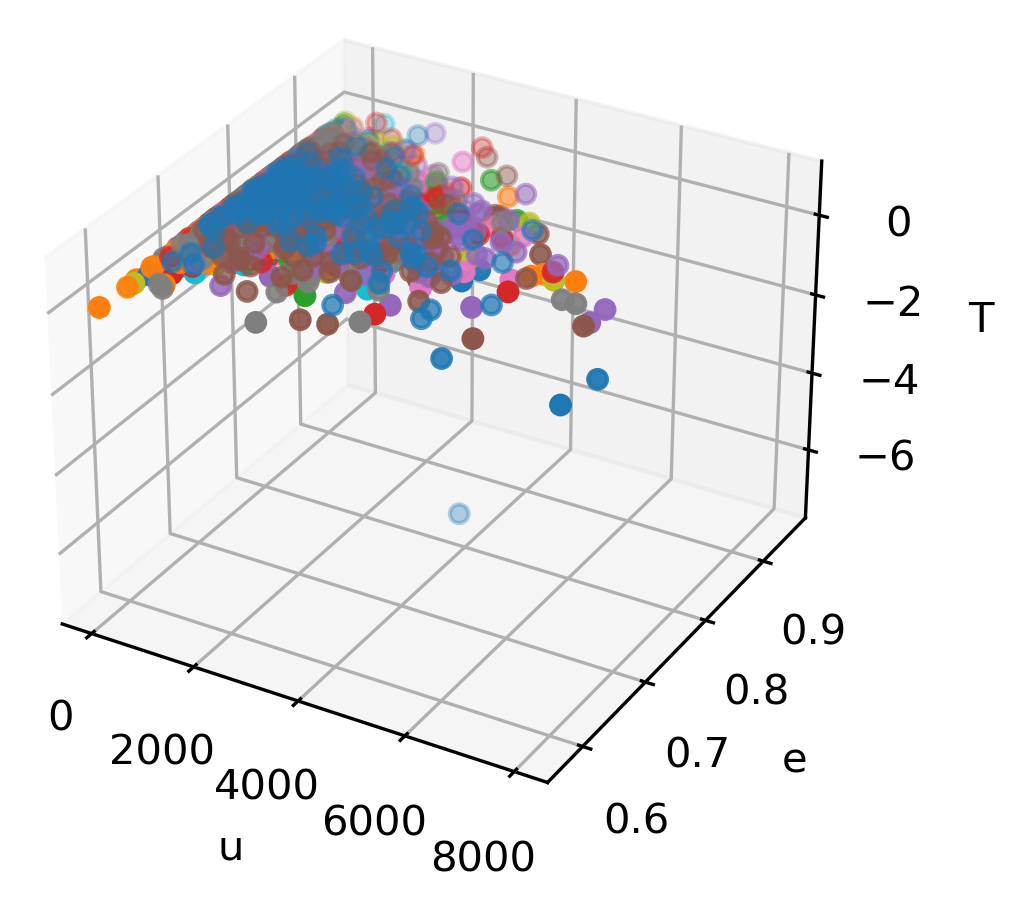}
       \label{fig:scatter_control}}
    \hfill
    \subfloat[Scatter\_e\_u\_T\_autism]{\label{fig:b}
       \includegraphics[width=0.45\linewidth]{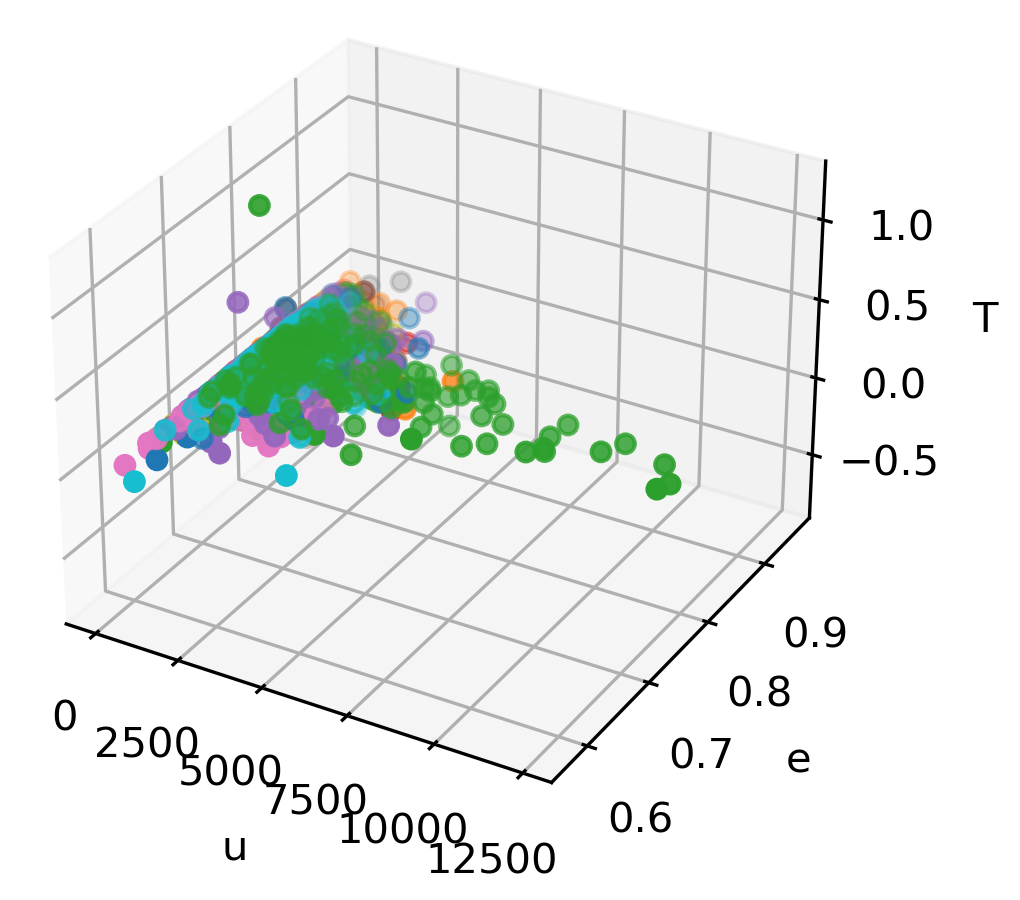}
       \label{fig:scatter_autism}}
    \caption{The 3-dimensional scatter plot of entropy, internal energy, and temperature at every time point, with each subject marked with a different color.}
    \label{3dscatter}
\end{figure*}

We can also examine the result by plotting the 3-dimensional scatter plot of entropy, internal energy, and temperature at every time point, as shown in Fig.\ref{3dscatter}. It is worth noticing that there are some outliers identifiable in the plot. These outliers are extreme values indicating the time points where transitions of brain states happen.

\begin{figure*}[h!]
    \centering
    \subfloat[Internal energy plot of subject $NYU\_0051038$]{\label{fig:a}
       \includegraphics[width=0.43\linewidth]{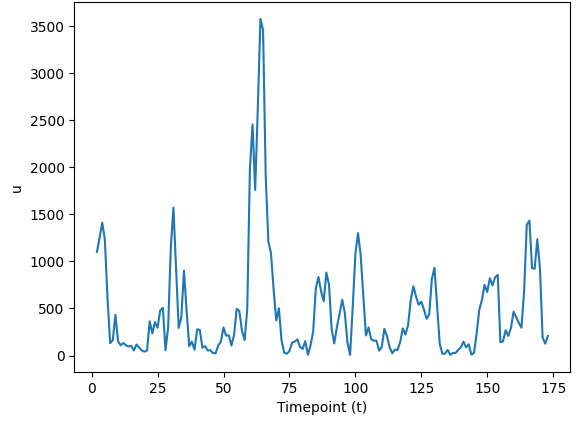}
       \label{fig:u_NYU_0051038}}
    \hfill
    \subfloat[Temperature plot of subject $NYU\_0051038$]{\label{fig:b}
       \includegraphics[width=0.45\linewidth]{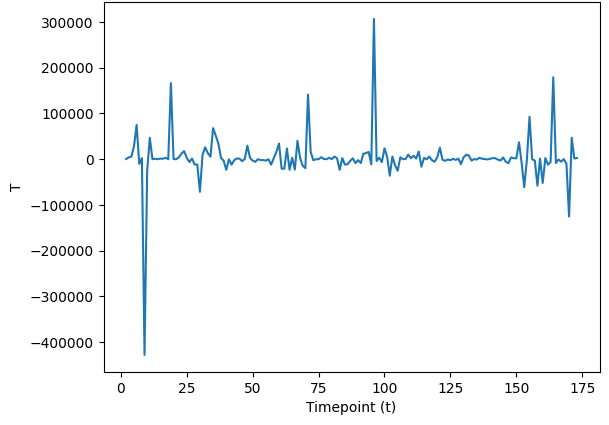}
       \label{fig:T_NYU_0051038}}\\
    \subfloat[Structural similarity plot of subject $NYU\_0051038$]{\label{fig:c}
       \includegraphics[width=0.45\linewidth]{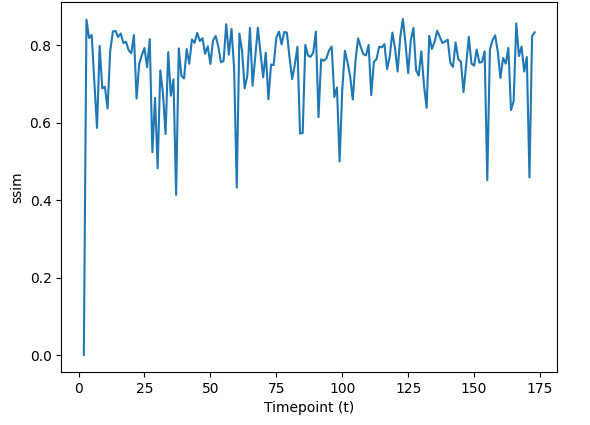}
       \label{fig:ssim_NYU_0051038}}
    \hfill
    \subfloat[Co-occurrence of T spikes and SSIM troughs of subject $NYU\_0050954$]{\label{fig:d}
       \includegraphics[width=0.43\linewidth]{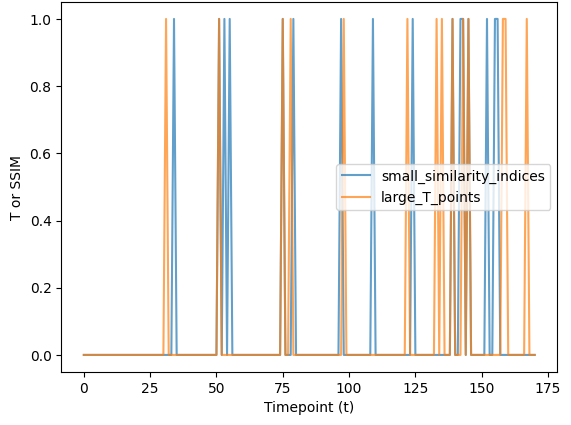}
       \label{fig:correspondence_NYU}}
    \caption{Illustration of time-varying physical quantities and brain state transition detection of subject $NYU\_0051038$}
    \label{NYU_0051038}
\end{figure*}

\subsection{Temperature spikes and brain state transitions}
By its definition and mathematical formulation, temperature symbolizes the structural change between the systems at $t$ and $t-1$ time instants. In brain network analysis, this implies critical events such as brain state transitions\cite{uddin2021brain}.

Using the subject $NYU\_0051038$ as an example, we show individual-level analysis of temperature spikes and their relation to brain state transition. The internal energy and temperature of $NYU\_0051038$ are illustrated in Fig. \ref{NYU_0051038}. Troughs in the SSIM plot suggest low similarities between the brain patterns in the past two time points, hence identifying changes in brain state. We show here that the T spikes and SSIM troughs of subject $NYU\_0051038$ correspond very well, motivating the following analysis of the statistical relationship between these two phenomena.

\subsection{Statistical testing}
To generalize the finding of the correspondence between T spikes and SSIM troughs of subject $NYU\_0051038$ to the group level, we test the MCC correlation between the occurrence of large temperature points and low structural similarity values and run a permutation test to determine the likelihood of obtaining a correlation value that is equally or more extreme. We summarize the average MCC and p-value of autism and control groups in Table. \ref{tab: MCC and p values}.

Despite the Matthews Correlation Coefficient being relatively low (around 0.25), the p-value of around 0.02 indicates that the probability of obtaining an MCC of 0.25 or greater, under the null hypothesis, is around 2\%. This signifies that this correlation is statistically significant. Therefore, we conclude that there is a statistically significant association between the two binary vectors, although the strength of this association is weak. This finding implies that the two binary vectors are more related than would be expected under the null hypothesis of no association.

\begin{table}[]
    \centering
    \begin{tabular}{c|c|c}
            & $MCC_{avg}$ & $p-value_{avg}$ \\
        \hline
        Autism & 0.23754078239629303 & 0.02542857142857143\\
        \hline
        Control & 0.2515742691126463 & 0.022375
    \end{tabular}
    \caption{MCC and p-value for autism and control groups}
    \label{tab: MCC and p values}
\end{table}

\section{Node energy observations and group differences}
\begin{figure}[h!]
    \centering
    \subfloat[Nodewise internal energy of autism group]{
        \includegraphics[width=0.85\linewidth]{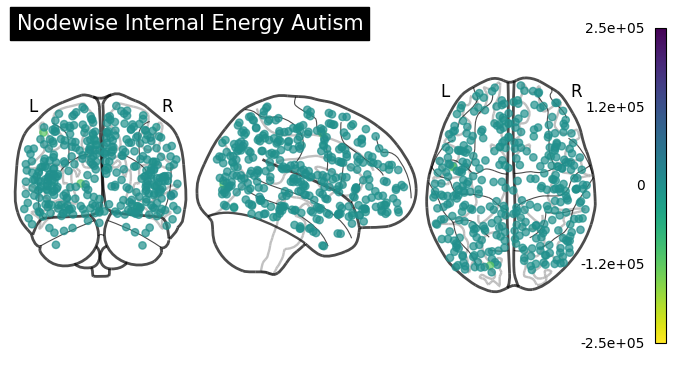}
        \label{fig:node_ui_autism}}\\
    \subfloat[Nodewise internal energy of control group]{
        \includegraphics[width=0.85\linewidth]{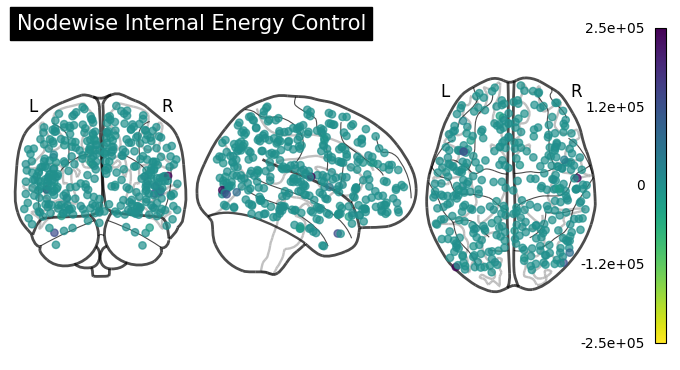}
        \label{fig:node_ui_control}}\\
    \subfloat[Group level differences of node internal energy]{
        \includegraphics[width=0.85\linewidth]{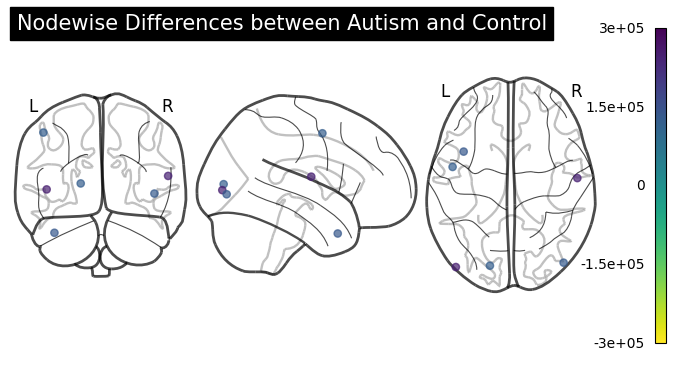}
        \label{fig:node_diff_group_level}}
    \caption{Illustration of time-varying physical quantities and brain state transition detection of subject $NYU\_0051038$}
    \label{node internal energy}
\end{figure}

We compute the groups' node energy for each node, as visualized in Fig. \ref{node internal energy}. In our experiments, since each node represents a corresponding brain region, the node energy is thus an important quantity, reflecting the activity level of that brain region within the network. We also calculate the differences between autism and control subjects and found significant discrepancies for several regions between them. We illustrate the nodes with the most prominent differences in Fig. \ref{node internal energy}(c), where drastic differences are found for nodes 4, 11, 20, and 200. We identified the corresponding brain regions and found relevant biological studies confirming that abnormalities in these regions are characteristic of autism patients. According to the Glasser atlas labels, we find the corresponding brain regions of these nodes where huge differences between control and autism occur, with the region names and their primary functions listed in Table. \ref{brain region labels}. 

\begin{table*}[h!]
    \centering
    \begin{tabular}{c|c|c|c}
        Index&Label&Brain region&Function\\
        \hline
        4 & Right\_V2 & secondary visual cortex & visual information processing \\ 
        \hline
        11 & Right\_PEF & parietal eye field & dorsal attention and eye movement \\ 
        \hline
        20 & Right\_LO1 & lateral occipital area 1 & visual perception (object recognition, motion detection) \\ 
        \hline
        130 & Right\_STSvp & superior temporal sulcus & default mode network, rest activities, auditory and social perception \\ 
        \hline
        200 & Left\_LO1 & lateral occipital area 1 & visual perception (object recognition, motion detection) \\ 
        \hline
        279 & Left\_43 & primary gustatory cortex & somatosensory, taste, and mouth functions.
    \end{tabular}
    \caption{Brain regions with most prominent differences and their corresponding functions.}
    \label{brain region labels}
\end{table*}

According to Table. \ref{brain region labels}, indices 4, 11, 20, and 200 correspond to regions for visual functions, indicating altered visual-related brain signals in autistic subjects. Such differences in visual information processing mechanisms between autism and control groups provide statistical evidence for former studies \cite{jones2011multimodal}, which found that autism patients have defective perceptions of other people's emotions and facial expressions, resulting in deteriorated levels of empathy and social ability. Region 130 is also about social perceptions and language, and it is a part of the default mode network (DMN), which is active during rest and involved in self-referential thinking, mind-wandering, and autobiographical memory. Differences related to region 130 could be linked to self-indulgence and higher information density produced by autistic brains at rest \cite{perez2013information}. Studies have also shown altered language development processes among autism patients \cite{boucher2003language}. There are also studies showing altered DMN functions and their relation with social withdrawal in autism \cite{padmanabhan2017default}, as well as atypical linkage mechanisms among visual, motor, and DMN regions \cite{washington2014dysmaturation}. All these findings provide explanations for our studies and make them more grounded. This real data demonstrates that our designed node energy effectively serves as a node feature to represent the network.

\section{Conclusions and future work}
In this paper, we introduced a novel thermodynamic framework to analyze the characteristics of graph representation in dynamic complex networks, specifically focusing on the application to brain network analysis in autism dataset. Our main methodology contribution, the Spectral Core Entropy (SCE) and Node Energy, allowed for a quantifiable representation of network characteristics that capture time-varying structural changes effectively. In our framework, the brain is treated as a thermodynamic system. It is found that its temperature and its node energy are both good representations of brain activities.

Our experimental findings validate the proposed framework by temperature analysis which reveals that spikes in temperature correspond to critical brain state transitions. This is confirmed by significant statistical correlations with structural changes in brain network connectivity. Moreover, we demonstrate significant differences in node energy distributions between autism and control groups, especially in regions linked to visual functions and the default mode network (DMN). These differences align with existing biological studies that associate such abnormalities with characteristic features of autism, including impaired emotion perception, social abilities, and language development. The application of this framework provided new insights into the intricate dynamics of brain states, highlighting its potential as a powerful tool in neuroscientific research and beyond. 

%We find the idea of viewing complex networks as thermodynamic systems, particularly the concept of treating nodes as particles within these systems, to be very intriguing. %
In the future, we plan to further develop the idea of viewing complex networks as thermodynamic systems by including more sophisticated thermodynamic models. For instance, we might examine whether node energy, which represents node degree, conforms to specific thermodynamic distributions such as the Boltzmann distribution. If networks conforming to a certain distribution demonstrate greater stability and representativeness, this could be applied to graph pruning. Furthermore, we intend to apply our thermodynamics analysis method to different datasets, such as stock and cryptocurrency datasets, to validate its capability in capturing the network characteristics and structures of other types of complex networks. By continuing to refine and expand our thermodynamic approach, we hope to provide a deeper understanding of the complex networks characteristics and to develop practical applications in the field of network science. 
%We believe that expanding physics parameters and principles holds great promise for both theoretical advancements and practical applications in the field of network science.

\section{Statement}
This work has been submitted to the IEEE for possible publication. Copyright may be transferred without notice, after which this version may no longer be accessible.

\bibliography{main}
\end{CJK}
\end{document}